\documentclass[11pt]{article}
\usepackage{graphicx}

\RequirePackage{xspace}

\newcommand{\BABARPubYear}    {03}

\newcommand{\BABARConfNumber} {024}
\newcommand{\SLACPubNumber} {10112}

\input pubboard/babarsym

\newcommand{\calB}{\mbox{${\cal B}$}}

\newcommand{\etapphi}{\mbox{$B^0\ra\eta^\prime \phi$}}

\newcommand{\DE}{\ensuremath{\Delta E}}
\newcommand{\mb}{\ensuremath{M_{ES}}}

\newcommand{\xf}{\mbox{${\cal F}$}}

\newcommand{\pepii}{\mbox{PEP-II}}
\newcommand{\UfourS}{\mbox{$\Upsilon(4S)$}}

\newcommand{\pvec}{{\bf p}}
\newcommand{\half}{\mbox{${1\over2}$}}

\def\babar{{\em B}{\footnotesize\em A}{\em B}{\footnotesize\em AR}}

\newcommand{\Betapphi}{\ensuremath{\calB(\etapphi)}}

\newcommand{\etal}{{\em et al.}}

\def\babar{\mbox{\slshape B\kern-0.37em{\relsize{-2} A}\kern-0.04em
    B\kern-0.37em{\relsize{-2} A\kern-0.04em R}}\xspace}

\def\babarbold{\mbox{\slshape B\kern-0.3em{\relsize{-2} A}\kern-0.1em
    B\kern-0.3em{\relsize{-2} A\kern-0.1em R}}\xspace}

\long\def\inst#1{\par\nobreak\kern 4pt\nobreak
    {\it #1}\par\vskip 10pt plus 3pt minus 3pt}

\begin{document}
{\pagestyle{empty}


\begin{flushright}
\babar-CONF-\BABARPubYear/\BABARConfNumber \\
SLAC-PUB-\SLACPubNumber \\
August 2003 \\
\end{flushright}

\par\vskip 3cm

\begin{center}
\Large \bf Search for  the {\boldmath $B$} Meson Decay to
\boldmath{$\eta^{\prime}$}   \boldmath{$\phi$}
\end{center}
\bigskip

\begin{center}
\large The \babar\ Collaboration\\
\mbox{ }\\
\today
\end{center}
\bigskip \bigskip

\begin{center}
\large \bf Abstract
\end{center}
We present preliminary results of a search for the decay $B^0\to
\eta^{\prime}\phi$. The data were recorded with the \babar\ detector
at the \pepii\ asymmetric-energy $B$-meson
Factory at SLAC and correspond to $89 $ million \BB\ pairs produced in
\epem\ annihilation at the \UfourS\ resonance. We find no
evidence for a signal and set  a  90\% CL upper limit of $\Betapphi < 1.0 
\times 10^{-6}$.

\vfill
\begin{center}
Submitted to the 
XXI$^{\rm st}$ International Symposium on Lepton and Photon Interactions at High~Energies, 8/11--8/16/2003, Fermilab, Illinois USA
\end{center}

\vspace{1.0cm}
\begin{center}
{\em Stanford Linear Accelerator Center, Stanford University, 
Stanford, CA 94309} \\ \vspace{0.1cm}\hrule\vspace{0.1cm}
Work supported in part by Department of Energy contract DE-AC03-76SF00515.
\end{center}

\newpage
} 

\begin{center}
\small

The \babar\ Collaboration,
\bigskip

%
B.~Aubert,
R.~Barate,
D.~Boutigny,
J.-M.~Gaillard,
A.~Hicheur,
Y.~Karyotakis,
J.~P.~Lees,
P.~Robbe,
V.~Tisserand,
A.~Zghiche
\inst{Laboratoire de Physique des Particules, F-74941 Annecy-le-Vieux, France }
% ...

\end{center}\newpage


\section{Introduction}
\begin{figure}[h]
\hspace{3cm}
\mbox{
\includegraphics[bb=85 155 535 605 ,angle=0,scale=0.35]{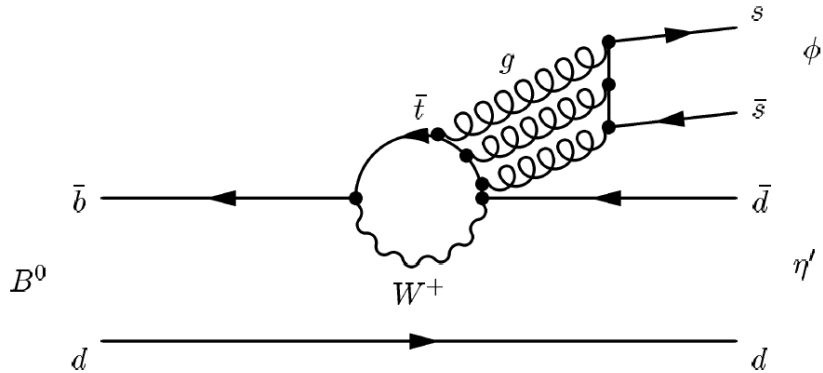}
}
\vspace{0.5cm}
\caption{A gluonic penguin diagram for $B^0\to\eta'\phi$.}
\label{fig:penguin}
\end{figure}

We present the results of a search  for 
the charmless $B$ meson decay $\B^0\to\eta^{\prime} \phi$. This $B$
decay mode to a pseudoscalar and a vector mesons is dominated by
penguin contributions. A gluonic penguin diagram is shown in     
Fig.~\ref{fig:penguin}. The branching ratio of this 
decay mode  is expected to be very small (in the range of $10^{-9}$ --
$10^{-7}$) \cite{Chen}. The study of 
this decay is  pertinent to the factorization model for nonleptonic decays and to the penguin mechanism. 

\vspace{-1cm}

\vspace{0.5cm}
        
CLEO, studying this decay with a sample of $3.3 \times 10^6$ \BB\
 pairs, found no evidence for a signal and set a 90\% CL
 upper limit of $\Betapphi < 31 \times 10^{-6}$ \cite{CLEO}.

\section{The \babarbold\ Detector and Data} \label{sec:detector}

The results presented in this paper are based on data collected
in 1999--2002 with the \babar\ detector~\cite{ref:babar}
at the PEP-II asymmetric $e^+e^-$ collider located  at the Stanford Linear Accelerator Center.  An integrated
luminosity of 81.9~fb$^{-1}$, corresponding to 
89 million \BB\ pairs, was collected at the $\Upsilon (4S)$
resonance
(``on-resonance'', center-of-mass energy $\sqrt{s}=10.58\ \gev$).
An additional 9.6~fb$^{-1}$ were collected about 0.040~GeV below
this energy (``off-resonance'') for the study of continuum background.
The asymmetric beam configuration in the laboratory frame
provides a boost of $\beta\gamma \approx 0.56$ to the $\Upsilon(4S)$,
increasing the momentum range of the $B$-meson decay products
up to $4.4\ \gevc$.
Charged particles are detected and their momenta measured by a
combination of a silicon vertex tracker (SVT), consisting of five layers
of double-sided detectors, and a 40-layer central drift chamber,
both operating in the 1.5~T magnetic field of a solenoid. Photons and
electrons are detected by a CsI(Tl) electromagnetic calorimeter (EMC).

Charged-particle identification (PID) is provided by the average 
energy loss (\dedx ) in the tracking devices  and
by an internally reflecting ring-imaging 
Cherenkov detector (DIRC) covering the central region. 
A Cherenkov-angle $\pi / K$ separation of better than 4$\sigma$  is 
achieved for tracks below $3\ \gevc$, decreasing to 
$2.4\,\sigma$ at the highest momenta in the final states  available in 
$B$-meson decays.

\section{Event Selection} 
\label{sec:presel}

The $B$-meson decay $\B^0\to\eta^{\prime} \phi$ is fully reconstructed
through the observation of $\etapr\ra\rho\gamma$ or
$\etapr\ra\eta\pi^+\pi^-$, where $\eta\ra\gamma\gamma$ and
$\phi\rightarrow K^+ K^-$.
   
Monte Carlo (MC) simulation \cite{GEANT}  of the decay mode under study and
of the continuum and \BB\ backgrounds is used to establish the event-selection
criteria.  The selection is designed to achieve high efficiency and
retain sidebands sufficient to characterize the background for
subsequent fitting.  Photons are required to have an  energy exceeding a 
threshold
that depends on the mode-dependent combinatorial background of the specific mode: an
$E_\gamma>0.2$ GeV for $\etaprrg$ candidates and $E_\gamma>$ 0.05 GeV 
for \etaprepp, $\eta\ra\gamma\gamma$.

We select $\eta^{\prime}$,$\eta$ , and $\rho^0$ candidates with the
following requirements on the invariant masses (in~\gevcc ) of their final
states: $0.920 < m_{\eta^{\prime}} <0.990$, $0.490 <  m_{\eta} < 0.600$, 
and $0.500 < m_{\rho^0} < 0.995$.
Tracks in $\eta^{\prime}$ ($\phi$) candidates must have DIRC, \dedx , and EMC
responses consistent with the pion (kaon) hypothesis.

A $B$-meson candidate is characterized kinematically by the
energy-substituted mass
$\mes = \sqrt{(\half s + \pvec_0\cdot \pvec_B)^2/E_0^2 - p_B^2}$ and the missing energy  $\DE = E_B^*-\half\sqrt{s}$, where the subscripts $0$ and
$B$ refer to the initial \UfourS\ and the $B$ candidate, respectively,
and the asterisk denotes the \UfourS\ rest frame. We require $|\DE|\le0.2$ GeV 
and
$5.2\le\mes\le5.29\ \gevcc$ .  The resolutions on these quantities are
about 0.030 GeV and $0.003\ \gevcc$, respectively.

To reject light-quark continuum background, we make use
of the angle $\theta_T$ between the thrust axes of the $B$ candidate and
the rest of the tracks and neutral clusters in the event (ROE), calculated in
the center-of-mass frame.  The distribution of $|\cos{\theta_T}|$ is
sharply peaked near $ 1$ for combinations drawn from the jet-like $q\bar q$
events, and nearly uniform for the  $B$ meson decays. 
We apply the cut 
$|\cos{\theta_T}|<0.9$.  A second $B$ candidate satisfying the
selection criteria occurs in about 7\% (18\%) of the events in the
decay mode $\etaprrg$ ($\etaprepp$). In this case we pick the combination with the 
$\eta^{\prime}$ mass closer to the PDG value~\cite{PDG2002}.

To discriminate against tau-pair and two-photon backgrounds we require
the event to contain at least five charged tracks.  

The remaining continuum background dominates the samples and is modeled
from sideband data in the maximum-likelihood (ML) fit described in 
section~\ref{sec:mlfit}. 

We use MC simulation of \BzBzb\ and \BpBm\ pair production and decay to
look for possible \BB\ backgrounds.  
Other charmless \B\ decays turn out to be the most likely other 
source of backgrounds in this study. To evaluate this source, 
we have produced a high statistics enriched sample of such \B\  
decays, using available estimates of the branching ratios.
From these studies we find no evidence of a
significant \BB\ background.

\section{Maximum Likelihood Fit}\label{sec:mlfit}

We use an unbinned, multivariate maximum likelihood fit to extract
the signal yield. 
With the cuts described in Section~\ref{sec:presel}, 
$B$-meson candidates are selected to 
match the kinematic structure of the decay mode $\B^0\to\eta^{\prime} \phi$.

\subsection{The Likelihood Function} \label{sec:like}

The likelihood function incorporates several weakly correlated variables. 
For the kinematics of the
$B$ decay  we use $\DE$ and $\mes$. We also include the mass of the 
$\eta^{\prime}$
 and , for the $B$ production and
energy flow a Fisher discriminant \xf . We combine the following four 
variables into 
a  Fisher discriminant: the angles with respect to the beam axis in the center-of-mass 
frame of the $B$ momentum and the $B$ thrust axis, and
the zeroth and second Legendre moments of the tracks and neutrals not used in reconstructing the decay $\eta^{\prime} \phi$
computed with respect to the $B$-candidate  thrust axis.

Thus, the input variables for the
$\eta^{\prime}_{\eta\pi\pi}$ channel 
are \DE , \mes , $m_\eta^{\prime}$, \xf\ , and the 
angular variable
$\mathcal{H}_{\phi}$.
In the analysis of the $\eta^{\prime}_{\rho\gamma}$ channel we add 
the angular variable $\mathcal{H}_{\rho}$.
The angular variable $\mathcal{H}_{\phi}$ is the cosine 
of the angle between the direction of 
the daughter $K^+$ with respect to 
the direction of the parent $B$ in the
$\phi$ rest frame. The angular variable 
$\mathcal{H}_{\rho}$ 
is defined as the cosine of the angle between the
direction of a $\rho^0$'s daughter 
and the direction of the parent $\eta^\prime$ in
$\rho^0$ rest frame.
These variables have a $\sin^2 \theta$ and $\cos^2 \theta$ shape in signal events, respectively.
In the continuum $q \bar{q}$ background the distributions of both angular 
variables are flat.

Since we measure the correlations among the observables in data to
be small, we take the
PDF for each event to be a product of the PDFs for the separate
observables.  We define two hypotheses $j$, where $j$ can be either signal or
continuum background.  The product PDF (to be evaluated with the
observable set for event $i$) is then given (for $\eta^{\prime}_{\eta\pi\pi}$) by:

\begin{equation}
{\cal P}^i_{j} =  {\cal P}_j (\mes) \cdot {\cal  P}_j (\DE) \cdot
 { \cal P}_j(\xf) \cdot {\cal P}_j (m_{\eta^{\prime}}) \cdot {\cal P}_j (\mathcal{H}_{\phi}) .
\end{equation}

The extended likelihood function for all input events $N$  is:

\begin{equation}
{\cal L} = \frac{\exp{(-\sum_j N_j)}}{N!} \prod_i^{N}  \sum_j N_j{\cal P}_j^i  
\end{equation}

where $N_j$ is the number of events of species $j$ to be found by the  fitter.

\subsection{Preparation of Inputs}

The PDF determination for the likelihood fit is accomplished with use of
Monte Carlo for the signal and on-resonance sideband
data for the continuum background.  

Peaking distributions (signal
masses, \DE\  and \xf) are parameterized as Gaussian functions. To obtain good fits 
to these samples we employ a sum of two or three Gaussian functions
  or Gaussian functions  
with different widths above and below the central value.  
Slowly varying distributions (combinatoric background under mass
or energy peaks) have linear behavior.  The combinatoric background in \mes\ is described by an empirical
phase-space function \cite{argus}.
$\mathcal{H}_{\phi}$ and $\mathcal{H}_{\rho}$ PDFs for  signal
are described by second-order polynomial (without the linear term)
while those for the background are described by a
linear and a second-order polynomial, respectively.

Control samples of $B$ decays to charm final states of similar topology
are used to account for the fidelity of the MC for variables describing
$B$-decay kinematics. We adjust the MC resolutions and central values, 
when necessary, by comparing data and simulation in these control samples.

\section{Systematics}
\label{sec:syst}

The uncertainty in the values of the parameters used in the PDFs are a source of systematic error.  
Variation of these parameters by 1 $\sigma$ 
induces an uncertainty of 0.03 signal events.  The fitting procedure itself is a source of 
systematic uncertainty.  Studies with simulated samples and background populations 
find that this uncertainty amounts to 0.21 signal events.



The remaining systematic uncertainties are multiplicative. Those due to MC statistics,
 track and photon multiplicities and  $B$ production are estimated from 
 auxiliary studies.  Published world
averages \cite{PDG2002}\ provide the $B$-daughter branching fraction
uncertainties. The systematic error due to all these sources is  estimated 
to equal  $0.06$ events.

The overall  systematic error is $0.2$ events (or $0.1\times 10^{-6}$
in terms of the branching fraction). This error is incorporated into 
the upper-limit calculation as described in the following section.



\section{Fit Results}

We find  no signal events in the on-resonance sample and for this
reason we adopt the Bayesian method to
calculate the 90\% CL upper limit. We perform ML fits in the
physical region of the parameters and integrate the likelihood from
zero to the branching fraction value where the integral
reaches 90\% of its asymptotic value.

\begin{figure}[h]
\begin{center}
\includegraphics[bb=85 155 535 605 ,angle=270,scale=0.35]{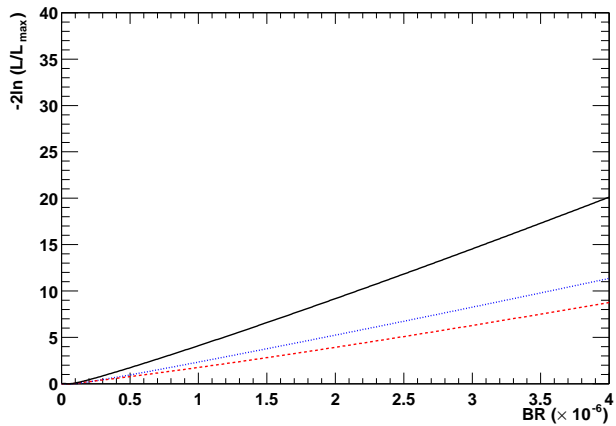}
\vspace{0.5cm}
 \caption{
Distributions of $-2\ln{\cal L}$ vs.~branching fraction for 
$\eta^{\prime} \phi$
decays.  Dotted, dashed and solid  lines correspond to
\etaprrg , $\eta'\to\eta\pi^+\pi^-$ and to the   combined decay modes,
respectively.}
\label{fig:chi2}
\end{center}
\end{figure}

The  $\eta^{\prime}_{\rho\gamma}\phi$ and
$\eta^{\prime}_{\eta\pi\pi}\phi$ modes are reconstructed with an
overall efficiency of 3.3\% and 2.1\%, respectively. The measured
value for the 90\% CL upper limit is $1.6 \times 10^{-6}$ for
$\eta^{\prime}_{\rho\gamma}\phi$ and $2.0 \times 10^{-6}$  for  
$\eta^{\prime}_{\eta\pi\pi}\phi$.
We combine these upper limit measurements  by forming for each
decay mode a  convolution of \calL\ 
from the fit with a Gaussian representing the uncorrelated systematic
error. The curves of $-2\ln{\calL}$
are shown in Fig. \ref{fig:chi2}, for both decay modes and for the sum of
the two.

\section{Conclusion}
\label{sec:conclusion}

We have performed a search for the charmless $B^0$ meson decay 
to $\eta^{\prime} \phi$. We find no signal and we set a 90\% CL upper limit:
\begin{eqnarray*}
\Betapphi & < & 1.0 \times 10^{-6}.
\end{eqnarray*}

\section{Acknowledgments}
\label{sec:Acknowledgments}

We are grateful for the 
extraordinary contributions of our \pep2\ colleagues in
achieving the excellent luminosity and machine conditions
that have made this work possible.
The success of this project also relies critically on the 
expertise and dedication of the computing organizations that 
support \babar.
The collaborating institutions wish to thank 
SLAC for its support and the kind hospitality extended to them. 
This work is supported by the
US Department of Energy
and National Science Foundation, the
Natural Sciences and Engineering Research Council (Canada),
Institute of High Energy Physics (China), the
Commissariat \`a l'Energie Atomique and
Institut National de Physique Nucl\'eaire et de Physique des Particules
(France), the
Bundesministerium f\"ur Bildung und Forschung and
Deutsche Forschungsgemeinschaft
(Germany), the
Istituto Nazionale di Fisica Nucleare (Italy),
the Foundation for Fundamental Research on Matter (The Netherlands),
the Research Council of Norway, the
Ministry of Science and Technology of the Russian Federation, and the
Particle Physics and Astronomy Research Council (United Kingdom). 
Individuals have received support from 
the A. P. Sloan Foundation, 
the Research Corporation,
and the Alexander von Humboldt Foundation.


\begin{thebibliography}{99}

\bibitem{Chen}
Y.~H.~Chen {\it et al.}, Phys.\ Rev.\ D {\bf 60}, 094014 (1999).

\bibitem{CLEO}
CLEO Collaboration,  Phys. Rev. Lett. {\bf 81}, 272 (1998).

\bibitem{ref:babar}
The \babar\ Collaboration, B.\ Aubert {\em et al.},
Nucl.\ Inst.\ Meth. A {\bf 479}, 1--116 (2002).

\bibitem{GEANT}
The \babar\ detector Monte Carlo simulation is based on GEANT : S.~Agostinelli {\it et al.}, 
Nucl.\ Instrum.\ Methods A {\bf 506}, 250--303 (2003).


\bibitem{argus}
With $x\equiv\mb/E_b$ and $\xi$ a parameter to be fitted, $f(x) \propto
x\sqrt{1-x^2}\exp{\left[-\xi(1-x^2)\right]}$.  See ARGUS Collaboration,
H.\ Albrecht \etal, \plb{241}, 278 (1990).

\bibitem{PDG2002}
Particle Data Group,
K.~Hagiwara {\em et al.}, Phys.\ Rev.\ D {\bf 66}, 010001 (2002).



\end{thebibliography}
\end{document}